\documentstyle[prb,aps,amstex,epsfig,eqsecnum]{revtex}

\begin{document}
\title{Hopping between Random Locations: Spectrum and Instanton}
\author{A. Zee $^{a}$ and Ian Affleck $^{a,b}$}
\address{
$^{a}$Institute for Theoretical Physics, University of California, \\
Santa Barbara,
CA 93106-4030, U.S.A. \\
$^{b}$Canadian Institute for Advanced Research and\\
Department of Physics and Astronomy, \\
University of British Columbia,\\
Vancouver, B.C., Canada, V6T 1Z1}
\maketitle

\begin{abstract}
Euclidean random matrices appear in a broad class of physical problems
involving disorder. The problem of determining their spectra can be mapped,
using the replica method, into the study of a scalar field theory with an
interaction of the type $e^{\psi ^{2}}$. We apply the instanton method to
study their spectral tails.
\end{abstract}

\section{Introduction}

Place $N$ points randomly in a $d$-dimensional Euclidean space of volume $V$%
.  Denote the locations of the points by $\vec{x}_{i}\left( i=1,...,N\right)
. $ Choose a suitably well-behaved function $f(\vec{x})$, vanishing as its
argument tends to infinity. Consider the $N\times N$ matrix
\begin{equation}
H_{ij}=f\left( \vec{x}_{i}-\vec{x}_{j}\right)  \label{ham}
\end{equation}
and solve the eigenvalue equation $\sum_{j}H_{ij}\psi _{j}=E\psi _{i}$.
We are interested in the density of eigenvalues $\rho (E)$ and the
localization properties of the eigenvalues $\psi _{i}$ as we average over
the ensemble of matrices [called Euclidean random matrices in Ref. ( %
\onlinecite{MPZ})] generated by placing $\vec{x}_{i}$ randomly. The limit $%
N\rightarrow \infty ,V\rightarrow \infty $, with the density $\rho \equiv {\
\frac{N}{V}}$ (not to be confused with $\rho (E)$ of course) held fixed, is
understood. In Ref. (\onlinecite{MPZ}) $\rho (E)$ was calculated in various
approximations. (A more involved version of (\ref{ham} ) was also studied.)
We will refer to this henceforth as model I.

This type of random matrix problem may be relevant to a broad class of
physical situations\cite{MPprl,Nie}, structural glasses\cite{MP0} and
amorphous semi-conductors for example.\cite{Elliott} Matrices of this type
have also appeared in the instantaneous normal mode analysis in the theory
of liquid\cite{liq} and in the bipartite matching problem\cite{Orland,MP} in
combinatorial optimization.

A venerable problem in the study of disordered systems is that of an
electron in a metal moving in a disordered array of impurities. The simplest
version of this problem ignores the periodic potential of the metal and
treats the impurities as $\delta $-function scatterers with Hamiltonian:
\[
H=-\nabla ^{2}+2\pi a\sum_{i=1}^N \delta (\vec{x}-\vec{x}_{i})
\]
where $\vec{x}_{i}$ are randomly located. The density of states in the tail
of the distribution, where it is exponentially small, was studied by
Lifshitz.\cite{Lifshitz} Later this calculation was reproduced and
extended using an instanton method\cite{FL} in the case of a repulsive
potential, $a>0$. A related problem of an electron moving in a white noise
Gaussian random potential was studied by Cardy.\cite{cardy} One of us\cite
{Affleck} extended this model to include a magnetic field $B$ in which case $%
\nabla ^{2}$ is replaced by $(\nabla -ieA)^{2}$ with $A$ the vector
potential. The density of eigenvalues in the tail of the distribution was
calculated. We will refer to this henceforth as model II.

In this note, we will elucidate the relation between model I and model II.
It turns out that the relationship only holds for the case of an attractive
potential, $a<0$. We will apply the instanton method to study the density of
eigenvalues $\rho (E)$ for large negative $E$. This calculation is similar
to that for $a>0$ already carried out earlier.\cite{FL} Finally, we extend
model I to include a magnetic field $B$ for $d=2$. Model I with a magnetic
field may be relevant for studying the quantum Hall transitions.

\section{Review of Field Theoretic Formulation}

In Ref. (\onlinecite{MPZ}) the problem was mapped into a quantum field
theory. We will review the procedure here.

We start with the replica identity
\begin{equation}
\Xi \,_{N}=\left\langle \frac{1}{det(z-H)^{n}}\right\rangle =\left\langle
e^{-ntr\log (z-H)}\right\rangle
\end{equation}
\begin{equation}
=\left\langle \int \prod_{i=1}^{N}\prod_{a=1}^{n}d\phi _{i}^{a}e^{
-\sum_{ija}\phi _{i}^{a*}\left( z\delta _{ij}-H_{ij}\right) \phi
_{j}^{a}}\right\rangle
\end{equation}
with $H_{ij}=f\left( x_{i},x_{j}\right) $. Here $a=1,2,...,n$ is the replica
index. We have used complex integration variables $\phi _{j}^{a}$ here
because eventually we want to study the problem with a magnetic field in
which case $H$ will be Hermitian rather than real symmetric as in (\ref{ham}
), for which real integration variables would have sufficed. Once we have
calculated $\Xi _{N}$ we can obtain the desired Green's function by
differentiation:
\begin{equation}
G(z)=\left\langle {\frac{1}{N}}tr{\frac{1}{z-H}}\right\rangle ={{\lim }
_{n\rightarrow 0}}\left( {\frac{1}{N}}\left( -{\frac{1}{n}}{\frac{\partial }{%
\partial z}}\right) \Xi _{N}\right) \
\end{equation}
Let us insert the representation of the identity (here $\psi $ and $\hat{\psi%
}$ denote two complex scalar fields)
\begin{equation}
1=\int D\hat{\psi}\delta \left( \hat{\psi}_{a}(x)-\sum _{i}\phi
_{i}^{a}\delta (x-x_{i})\right) =\int D\psi D\hat{\psi}e^{i\sum _{a}\int
dx\psi _{a}^{*}(x)\left( \hat{\psi}_{a}(x)-\sum_{ia}\phi _{i}^{a}\delta
(x-x_{i})\right) +h.c.}
\end{equation}
into the functional integral refining $\Xi $, thus obtaining

\begin{equation}
\Xi_{N} = \int D \hat{\psi} D \psi e^{\sum_{a} \int \int dx dy \hat \psi
_{a}^{*} (x) f (x,y) \hat{\psi}_{a} (y)} e^{i \int dx \psi_{a}^{*} (x)\hat %
\psi_{a}(x) + h.c.} \left\langle J \right\rangle
\end{equation}
where
\begin{equation}
J \equiv \int d \phi \ \ e^{- z \sum_{ia} | \phi_{i}^{a}|^{2} - i
\sum_{ia} \left( \phi_{i}^{a} \psi_{a}^{*} (x_{i}) + h.c. \right)}
\end{equation}
is a functional of $\psi$.

Integrating out $\phi $, we find
\begin{equation}
\left\langle J\right\rangle =\left\langle {\prod_{i}}\left( {\frac{2\pi }{z}}
\right) ^{n}e^{-{\frac{1}{z}}{\sum_{i}}|\psi
_{a}(x_{i})|^{2}}\right\rangle
\end{equation}
Recall that the average $\left\langle J\right\rangle $ means $\int {\frac{
dx_{1}}{V}}{\frac{dx_{2}}{V}}...{\frac{dx_{N}}{V}}J$ we see that the
multiple integral factorizes and we obtain
\begin{equation}
\left\langle J\right\rangle =A^{N}
\end{equation}
with
\begin{equation}
A\equiv {\frac{1}{V}}\int dxe^{-{\frac{1}{z}}{\sum_{a}}|\psi _{a}(x)|^{2}}
\end{equation}
We have taken the $n\rightarrow 0$ limit wherever we are allowed to do so.
The integral over $\hat{\psi}$ can be done immediately giving
\begin{equation}
\Xi _{N}=\int D\psi e^{\int \int dxdy\psi ^{*}(x)f^{-1}(x,y)\psi (y)}A^{N}
\end{equation}
To put the factor $A^{N}$ into the exponential we follow Gibbs and introduce
the grand canonical ensemble
\begin{equation}
Z(\alpha )\equiv {\sum_{N=0}^{\infty }}{\frac{\alpha ^{N}}{{N!}}}\Xi
_{N}=\int D\psi e^{{\int \int \psi ^{*}f^{-1}\psi +{\frac{\alpha }{V}}\int
dxe}^{^{-{\frac{1}{z}}{\sum_{a}}|\psi _{a}(x)|^{2}}}}{}  \label{grand}
\end{equation}

In other words, instead of focusing on the original problem of studying a
random $N$ by $N$ matrix we now consider an ensemble of such matrices with $N
$ varying over the non-negative integers. We expect the sum in (\ref{grand}
) to be dominated by some values of $N$:
\begin{equation}
\left\langle N\right\rangle ={\frac{\partial }{\partial \alpha }}\log
Z(\alpha )={\frac{\alpha }{V}}\left\langle \int dxe^{-{\frac{1}{z}}{\sum_{a}}|
\psi _{a}(x)|^{2}}\right\rangle @>>{n\rightarrow 0}>\alpha
\end{equation}
Defining the density of points as $\rho \equiv {\frac{\left\langle
N\right\rangle }{V}},$ we obtain $Z(\alpha )=\int D\psi e^{-S(\psi )}$ with
the action
\begin{equation}
S(\psi )=-\sum_{a}\int \int dxdy{\psi _{a}^{*}}(x)f^{-1}(x,y)\psi
_{a}(y)-\rho \int dx\ e^{-{\frac{1}{z}}\sum_{a}|\psi _{a}(x)|^{2}}
\end{equation}

The action $S(\psi )$ defines a non-local field theory. Up to this point,
any $f(x,y)$ could have been used. A particularly convenient choice is the
Yukawa function
\begin{equation}
f(x,y)=f(x-y)=(-)\int \frac{d^{d}k}{(2\pi )^{d}}\quad \frac{e^{ik(x-y)}}{
k^{2}+m^{2}}  \label{f}
\end{equation}
The overall minus sign is included so that the resulting field theory would
have the standard kinetic energy term:
\begin{equation}
S(\psi )=\int d^{d}x\left[ {\sum_{a=1}^{n}(}|\nabla \psi
_{a}|^{2}+m^{2}|\psi _{a}|^{2})-\rho e^{-{\frac{1}{{z}}}{}{\sum_{a=1}^{n}}
|\psi _{a}|^{2}}\right]  \label{field}
\end{equation}

One remark about the spectrum: The eigenvalue equation implies that for a
general $f(x)=\int \frac{d^{d}k}{(2\pi )^{d}}\tilde{f}(k)$
\begin{equation}
E=\sum_{ij}{\psi _{i}^{*}}H_{ij}\psi _{j}=\sum_{ij}{\psi _{i}^{*}}\int
\frac{d^{d}k}{(2\pi )^{d}}\tilde{f}(k)e^{ik(x_{i}-x_{j})}\psi _{j}=\int
\frac{d^{d}k}{(2\pi )^{d}}\tilde{f}(k)|\sum_{j}e^{-ikx_{j}}\psi _{j}|^{2}
\end{equation}
Thus, for $\tilde{f}(k)<0$, as is the case with the choice in (\ref{f}), the
eigenvalues are all negative.

A few remarks about renormalizability and dimensionless parameters  are in
order. This field theory is ultraviolet finite in $d=1$, renormalizable in $%
d=2$ (although requiring an infinite number of counter-terms in contrast to
 the sine Gordon theory with its special symmetry) and non-renormalizable
in $d>2$. The original problem, in model I, is well defined for any well
defined $f(x).$ However, for the particular choice of $f(x)=(-)\int \frac{%
d^{d}k}{(2\pi )^{d}}\quad \frac{e^{ikx}}{k^{2}+m^{2}}$ we used in order for
the field theory to have a normal kinetic energy term, $f(0)$ does not exist
for $d>1$. One way of regularizing is to write
\begin{equation}
f(x)=(-)\int \frac{d^{d}k}{(2\pi )^{d}}e^{ikx}(\frac{1}{k^{2}+m^{2}}-\frac{1%
}{k^{2}+M^{2}})=(-)\int \frac{d^{d}k}{(2\pi )^{d}}\frac{e^{ikx}}{%
(k^{2}+m^{2})(k^{2}+M^{2})}(M^{2}-m^{2}).
\end{equation}
The field theory would then be cut off correspondingly by the Pauli-Villars
mass $M.$ The ultraviolet difficulties in model II can be understood as
arising from the fact that a $\delta $-function potential is too singular to
have a well-defined spectrum for $d>1$. Physical ultra-violet
regularizations would replace the $\delta $-function potential by a smooth
function.

Thus, in $d=1$ the density of states should be a well-defined function of $E$
, $\rho $ and $m$. In $d=2$ it should be expressible in terms of
renormalized parameters but in $d>2$ it will be strongly dependent on the
details of the cut-off.

Ignoring cut-offs, we see that the field theory (\ref{field}) contains two
dimensionless quantities:
\[
\nu \equiv \rho /m^{d}
\]
and
\[
\Omega \equiv |E|m^{2}/\rho ,
\]
where we have set $z=-|E|$ in light of an earlier remark. The parameter $\nu
$ has the physical interpretation of the number of points in the correlation
volume of the function $f$. Consider expanding the action:
\begin{equation}
S(\psi )=\int d^{d}x{\sum_{a=1}^{n}(}|\nabla \psi _{a}|^{2}+(m^{2}-\frac{
\rho }{|E|})|\psi _{a}|^{2})-\frac{\rho }{|E|^{2}}({\sum_{a=1}^{n}}|\psi
_{a}|^{2})^{2}+\cdots ].
\end{equation}
The coupling constant of the $\psi ^{2n}$ term is of order $\rho /E^{n}$.
Since this has dimensions of $(\hbox{mass})^{d-n(d-2)}$, we see that the
condition for the dimensionless coupling constant to be small is (for $n\geq
2)$:
\begin{equation}
\Omega ^{n}\nu ^{n-1}>>1.  \label{condition}
\end{equation}
Here we have assumed $\Omega >1$, a condition neccessary for the
perturbative stability of the theory, as discussed in Sec. IV. If $\Omega $
is only slightly greater than 1 then we get  more complicated conditions:
\begin{equation}
\Omega ^{n}\nu ^{n-1}\left[ 1-{\frac{1}{\Omega }}\right] ^{d/2+n(1-d/2)}>>1.
\label{condition2}\end{equation}
We see that the conditions of Eq. (\ref{condition}) or (\ref{condition2})
  require at least one of
the two parameters $\Omega $ and $\nu $ to be large.

\section{Relationship between Models I and II}

Consider model II in the case of an attractive potential, $a<0$ and assume
that all points are far apart compared to the range of the ground state
wave-function, $\vec{\psi}_{0}(\vec{r})$, for a single $\delta $-function
potential. Then the lowest energy states will be formed by tunnelling
processes between these lowest bound states. (As mentioned above, for $d>1$,
the $\delta $-function potential must actually be replaced by some smooth
function.) To formalize this, we go to the tight binding approximation. The
hopping amplitude for an electron to go from a potential well at the origin
to a potential well a distance $\vec{R}$ away is given by the overlap
integral
\begin{equation}
t(\vec{R})=2\pi a\psi _{0}(\vec{0})\psi _{0}(\vec{R})
\end{equation}
where $\psi _{0}(\vec{r})$ is the solution of
\begin{equation}
-\nabla ^{2}\psi _{0}(\vec{r})=E\psi _{0}(\vec{r})
\end{equation}
(for $r>0$) with $E$ the (negative) binding energy in the single site
potential. The hopping amplitude $t(\vec{R})$ is thus just a constant times
a wave function $\psi _{0}(\vec{R})$ and so essentially equal to the Yukawa
function in Eq. (\ref{f}) with $|E|$ playing the role of $m^{2}.$

In ref.(\onlinecite{FL,Affleck}), it was shown that model II can be
represented in terms of a field theory with the action
\[
S(\psi )=\int d^{d}x\left[ {\sum_{a=1}^{n}(}|\nabla \psi _{a}|^{2}-E|\psi
_{a}|^{2})+\rho (1-e^{-{2\pi a}{}{\sum_{a=1}^{n}}|\psi
_{a}|^{2}})\right].
\]
In light of the preceding discussion, this is precisely what we would
expect. Thus, we see that model II corresponds to model I with a specific
choice of the function $f(\vec{x}),$ with the map of the complex variable $z$
to the inverse of the strength of the potential $2\pi a$ and $m^2$ to $-E$.

\section{Instanton Analysis}

In Ref. (\onlinecite{MPZ}) the density of eigenvalues was studied starting
with the field theory(\ref{field}). Here as promised we will use the
instanton method\cite{coleman} to study the tail of the spectrum.

In order for $G(z)$ to have an imaginary part the functional integral
defining the field theory has to be ``sick''. Otherwise, the functional
integral, if well defined, is manifestly real. Let us check this statement.
Consider the regime in which $z$ is real positive so that $z=|z|$. Then the
potential
\begin{equation}
V(\psi )=m^{2}|\psi |^{2}-\rho \left( e^{-{\frac{1}{|z|}}|\psi
|^{2}}-1\right) \simeq \left( m^{2}+{\frac{\rho }{|z|}}\right) |\psi
|^{2}+\ldots
\end{equation}
is well-behaved at large $|\psi |$and so perturbation theory should be fine.
Indeed, by the argument given earlier we expect all the eigenvalues to be
negative, and so $G(z)$ should not have an imaginary part.

In contrast, for $z$ negative, we have
\begin{equation}
V(\psi )=m^{2}|\psi |^{2}-\rho \left( e^{{\frac{1}{|z|}}|\psi
|^{2}}-1\right) \simeq \left( m^{2}-{\frac{\rho }{|z|}}\right) |\psi
|^{2}+\ldots  \label{negz}
\end{equation}
The potential is unbounded below for large $|\psi |$. Perturbation theory
fails and $G(z)$ could well have an imaginary part. From (\ref{negz}) we see
that we should distinguish two regimes: small eigenvalue $|z|\leq \rho
/m^{2} $ and large eigenvalue $|z|\geq \rho /m^{2}$. In the large eigenvalue
regime the potential $V(\psi )$ starts out concave upward before becoming
unbounded and so there can be an instanton configuration connecting $\psi =0$
to $\psi =\psi _{0}$ where $V(\psi _{0})=0$. In the small eigenvalue $%
|z|\leq \rho /m^{2}$ regime the potential $V(\psi )$ starts out downward and
the instanton approach does not apply. Some other non-perturbative method is
needed to study the spectrum.

To summarize, we have three regimes: (I) $z$ real positive, no eigenvalue
and $G(z)$ does not have an imaginary part, (IIa) $z$ real negative and $%
|z|\leq \rho /m^{2}$ (that is, $\Omega \leq 1)$ non-perturbative regime, and
finally (IIb) $z$ real negative and $|z|\geq \rho /m^{2}$ (that is, $\Omega
>1)$, and the instanton approach applies provided that the coupling
constants are small, the conditions of Eq. (\ref{condition}). In this paper
we focus on the regime (IIb) and hence study the tail of the eigenvalue
distribution.

We will now give a heuristic argument on how large negative eigenvalues can
occur. Suppose that in the distribution of the $N$ random points we have an
isolated cluster of $k$ points within a length scale comparable to the
length scale $l$ characteristic of $f(x)$ ($l=1/m$ in our specific example).
The Hamiltonian $H$ then contains a $k$ by $k$ block whose entries are of
order $f(0),$ thus giving us one eigenvalue of order $kf(0)$ (recall that $%
f(0)$ is negative in our example.) The probability $P_{k}$ of obtaining such
a cluster is given by the Poisson distribution $P_{k}\propto e^{-(\rho
l^{d})}\frac{(\rho l^{d})^{k}}{k!}$ and so heuristically we obtain the
estimate for the probability of obtaining a large negative eigenvalue $E$ to
be
\begin{equation}
P(E)\sim e^{-|\frac{E}{f(0)}|\log |\frac{E}{f(0)}|}  \label{heuristic}
\end{equation}
We expect that this heuristic argument will work best for $d=1.$

For $z=E$ large and negative, let us scale $\psi =\sqrt{|E|}\varphi $ and $%
x=y/m$ to rewrite the action as
\[
S(\varphi )=\frac{|E|}{m^{d-2}}\int d^{d}y{(}|\nabla _{y}\varphi
|^{2}+(|\varphi |^{2}-\frac{\rho }{|E|m^{2}}e^{|\varphi |^{2}})
\]
In terms of the dimensionless quantities $\Omega \equiv \frac{|E|m^{2}}{\rho
}$ and $\nu \equiv \rho /m^{d},$we see that
\[
e^{-S}\sim e^{-\nu \Omega h(\Omega )}
\]
for some function $h.$

When the conditions of Eq. (\ref{condition}) are satisfied, the functional
integral giving $Z(\alpha )$ is dominated by the extremum of the action $S$,
namely the instanton. We assume that the instanton is spherically symmetric,
so that $q\equiv |\varphi |$ is a function of $|\vec{y}|$ only. Following
standard practice in instanton analysis we identify $q$ as the position of a
particle along a line and $t\equiv \frac{1}{\sqrt{2}}|\vec{y}|$ as time, we
have the equation of motion
\begin{equation}
\frac{d^{2}q}{dt^{2}}\quad +\quad \frac{(d-1)}{t}\quad \frac{dq}{dt}\quad =-
\frac{dU}{dq}
\end{equation}
with the potential
\begin{equation}
U(q)=-q^{2}+{\frac{1}{\Omega }}\left( e^{q^{2}}-1\right)
\end{equation}
as shown in figure (1).

\begin{figure}
\epsfysize=5 cm
\centerline{\epsffile{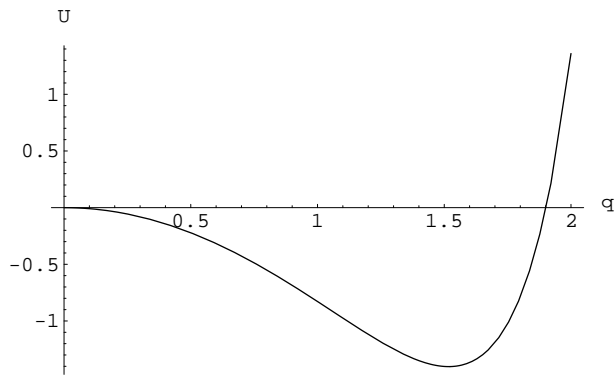}}
\caption{Potential for the single particle problem 
for $\Omega =10$.}
\label{fig:potential}
\end{figure}

For $d>1$, there is a time-dependent friction term. For $d=1$ there is no
friction.

Let us define $q_{0}$ and $q_{min}$by $U\left( q_{0}\right) =0$ and $%
U^{\prime }\left( q_{min}\right) =0$ respectively:
\begin{equation}
q_{min}^{2}=\log \Omega
\end{equation}
and
\[
q_{0}^{2}=\log \left( \Omega q_{0}^{2}+1\right)
\]
We have $q_{0}^{2}\simeq \log \Omega +\log \log \Omega +\cdots $ for $\Omega
>>1.$ We wish to determine the trajectory of the particle, starting at $%
q=q_{0}$ at $t=0$ and ending at $q=0$ at $t=\infty $. In the end, we would
like to calculate the action associated with the trajectory.

Clearly, we need to treat the cases $d=1$ and $d>1$ separately.

For $d=1$, we have explicitly $S=2\sqrt{2}|E|m\int_{0}^{\infty }dt\left(
\frac{1}{2}{\dot{q}}^{2}-U(q)\right) .$ It is convenient to work with the
reduced action $S_{r}=\int_{0}^{\infty }dt\left( \frac{1}{2}{\dot{q}}
^{2}-U(q)\right) $. With the initial conditions specified, the equation of
motion integrates to ${\dot{q}}^{2}=-2U(q)$ and so the trajectory is given
by
\begin{equation}
t=\int_{q}^{q_{0}}\quad {\frac{{dq^{\prime }}}{{\sqrt{-2U(q^{\prime )}}}}}
\label{traj}
\end{equation}
Evaluated for this trajectory (\ref{traj}) the action is equal to
\begin{equation}
S_{r}=\int_{0}^{\infty }dt{\dot{q}}^{2}=\int_{0}^{q_{0}}dq\sqrt{-2U(q)}
\end{equation}
which can be easily integrated numerically as a function of $\Omega .$

Our instanton analysis thus predicts that for $d=1,$ the tail of the
eigenvalue distribution should go like
\begin{mathletters}
\begin{equation}
P(E)\sim LF(|E|,\rho )e^{-2\sqrt{2}\nu \Omega S_r(\Omega )}  \label{inst}
\end{equation}
where $L$ denotes the size of the system (which we need to compare with
direct numerical diagonalizations of $H)$ and $F(|E|,\rho )$ the infamous
determinant factor in instanton calculations (which we have not computed.)
We expect $F(|E|,\rho )$ to be slowly varying compared to the exponential
factor $e^{-2\sqrt{2}\nu \Omega S_{r}(\Omega )}.$

We have done some numerical work in which we diagonalized the matrix $H$
directly for $N$ ranging up to 1000. In figure (2) we show the numerical data
for $N=1000,$ $\rho =1.2,$ and $m=(1.3)^{-1}.$ Only the tail of the
distribution, which we take to be comprised of the $250$ eigenvalues with the
most negative $E$, is displayed. We take (\ref{inst} ) and compute the
integrated number of eigenvalues $N(E)=N\int_{-\infty }^{E}dE^{\prime
}P(E^{\prime })$, treating $F(|E|,\rho )$ as a constant $C$ in the range of $%
E$ of interest. ($N(E)$ is not to be confused with $N$ of course; $N(\infty %
)=N.)$ We then do a one parameter fit in $C$ to the numerical data. As
expected, the theoretical curve appears to fit the numerical data in the
applicable regime, Eq. (\ref{condition}).  It should be kept in mind that the
theoretical curve fails to make sense as $\Omega$ approaches $1$ which, for our
particular parameter choice, corresponds to $E=-2.03$.  Also, the numerical
data is clearly dominated by finite size effects in the extreme tail where
$N(E)$ is O(1).

\begin{figure}
\epsfxsize=10 cm
\centerline{\epsffile{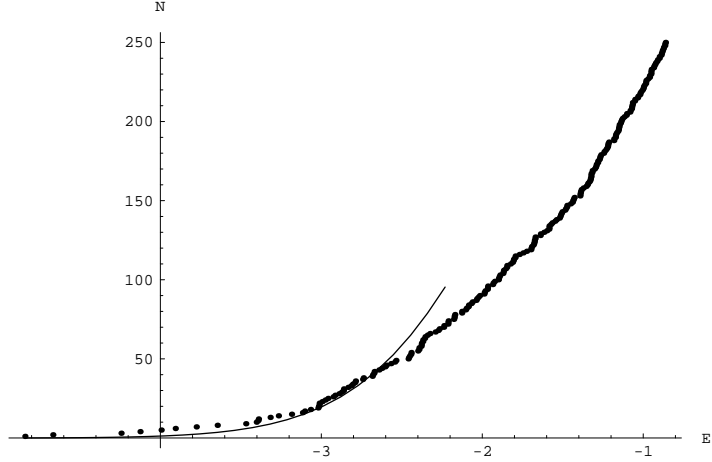}}
\caption{The number of eigenvalues with energy $<E$, plotted versus $E$.
Only the low energy tail of the distribution is plotted.}
\label{fig:density}
\end{figure}

Analytically, we are able to study the problem only for $\Omega >>1$ in
which case
\end{mathletters}
\begin{equation}
S_{r}\simeq \frac{1}{\sqrt{2}}\log \Omega   \label{actionlargeomega}
\end{equation}

We can check this result by studying the trajectory of course. Think of the
particle being released at $q_{0}$ with zero velocity. We divide the
trajectory into three regimes: (1) short time, during which the particle
moves rapidly from $q_{0}$ to $q_{min}$, (2) intermediate time, during which
the particle passes through the minimum of the potential at $q_{min}$, and
(3) long time, during which the particle slowly climbs the hill to $q=0$. A
priori, it is not immediately obvious whether the short or long time regime
gives the dominant contribution to $S_{r}$. A detailed calculation, which we
now outline, shows that the long time regime dominates for $\Omega \gg 1$.

In the short time regime, we approximate $U(q)\simeq 2q_{0}^{3}(q-q_{0})$,
thus obtaining $q(t)=q_{0}-\frac{1}{2}q_{0}^{3}t^{2}+\cdots .$ Similarly, we
approximate $U(q)$ in the other two regimes and match the trajectory, thus
obtaining in the intermediate time regime $q(t)=q_{min}-\frac{1}{\sqrt{2}}
\sin (\sqrt{2\log \Omega }(t-t_{min}))+\cdots $ and in the long time regime $%
q(t)=q_{min}e^{-(t-t_{min})}+\cdots $ where $t_{min}=(\frac{2}{q_{0}^{3}}
(q_{0}-q_{min}))^{\frac{1}{2}}\rightarrow \frac{(2\log \log \Omega )^{\frac{%
1 }{2}}}{\log \Omega }.$ Substituting into $S_{r},$ we find that $S_{r}$
goes like $O((\log \log \Omega )^{\frac{3}{2}}),$ $O((\log \log \Omega )^{%
\frac{1 }{2}})$, and $\frac{1}{\sqrt{2}}\log \Omega $ in the short,
intermediate, and long time regime respectively. Thus, the long time regime
dominates and we obtain
\begin{equation}
S\simeq 2|E|m\log \frac{|E|m^{2}}{\rho }  \label{S1D}
\end{equation}
Using $f(0)=-1/2m$, in $d=1$, from Eq. (\ref{f}), we see that the instanton
result, Eq. (\ref{S1D}), is in agreement with the heuristic argument of Eq.
( \ref{heuristic}).

The limit $\Omega =1+\delta ,$with $\delta \ll 1$ is also interesting. We
have $q_{0}\simeq \sqrt{2\delta }$ and $q_{min}\simeq \sqrt{\delta },$ with $%
U(q)\simeq -\delta q^{2}+\frac{1}{2}q^{4}.$ The trajectory is explicitly
found to be $q(t)=\sqrt{2\delta (1-\tanh \sqrt{2\delta }t)}.$ We obtain
\begin{equation}
S\simeq \frac{3}{2\sqrt{2}}(\frac{|E|m^{2}}{\rho }-1)^{\frac{3}{2}}
\end{equation}

We now turn to higher dimensions $d=2$ or $3.$ The equation of motion now
includes a friction term
\[
\frac{d^{2}q}{dt^{2}}\quad +\quad \frac{(d-1)}{t}\quad \frac{dq}{dt}\quad
=q(1-\frac{1}{\Omega }e^{q^{2}})
\]
At small time, near $q_{0}$ we have $\frac{d^{2}q}{dt^{2}}\quad +\quad \frac{
(d-1)}{t}\quad \frac{dq}{dt}\simeq -\gamma \quad $ and the solution $%
q(t)=q_{0}-\frac{\gamma }{2d}t^{2}+\cdots .$ Henceforth we will consider
only the large $\Omega $ limit. The minimum of the potential at $q_{min}$ is
reached at time $t_{min}=\sqrt{2d(q_{0}-q_{min})/\gamma }\simeq \sqrt{d\frac{
\log \log \Omega }{(\log \Omega )^{2}}}\rightarrow 0$ for large $\Omega .$
Focussing then on the large time regime we can approximate the equation of
motion by
\[
\frac{d^{2}q}{dt^{2}}\quad +\quad \frac{(d-1)}{t}\quad \frac{dq}{dt}\quad
\simeq q
\]
Remarkably, for $d=3,$ we have an exact solution $q(t)=\frac{c}{t}e^{-t}$
with constant $c$ fixed by matching to the small time solution so that $%
q(t)=q_{min}t_{min}\frac{1}{t}e^{-(t-t_{min})}.$ Putting this solution into
the action we find
\[
S\sim \frac{(\log \Omega )^{3}}{\log \log \Omega }
\]
for large $\Omega .$

For $d=2,$ we mention for what it is worth that the equation of motion has
an exact solution for large time in terms of a Bessel function:
\begin{equation}
q(t)=c^{\prime }K_{0}(t) @>>{t\rightarrow \infty}> c^{\prime }\sqrt{\frac{%
\pi }{2t}}e^{-t}(1-\frac{1}{8t}+\cdots )
\end{equation}

In fact, this instanton calculation seems to agree with the heuristic
argument, given earlier in this section, only in $d=1$. While we do not
 understand this discrepancy, we suspect that it is related to the
existence of ultraviolet divergences for $d>1$. Note that the heuristic
estimate of Eq. (\ref{heuristic}) involves $f(0)$, which is ultraviolet
divergent for $d>1$.  On the other hand, the instanton result appears 
to be ultraviolet finite.  However, this is somewhat illusory, since
we expect loop corrections to bring in divergences, for $d>1$.
This issue requires further investigation.

\section{Magnetic Field}

In this paper, we propose a natural generalization of model I to include a
magnetic field. In principle, this would define a new model for studying
localization properties in the quantum Hall system.

There are two possibilities. In the first, we write
\begin{equation}
f\left( \vec{x_{i}},\vec{x_{j}}\right) =g\left( \vec{x_{i}},\vec{x_{j}}
\right) e^{i\int_{x_{i}}^{x_{j}}\vec{A}.d\vec{l}}  \label{magf}
\end{equation}
To have a well defined model, we have to specify the path joining $\vec{%
x_{i} }$ to $\vec{x_{j}}$. The simplest choice is to take a straight line,
then the phase factor in (\ref{magf})becomes $e^{{\frac{1}{2}}iB\left( \vec{%
x_{i}} \times \vec{x_{j}}\right) }$for a constant magnetic field $B$.

Unfortunately, $f^{-1}(x,y)$, the functional inverse of $f(x,y)$ in (\ref
{magf}), does not have a particularly simple form, and so the corresponding
field theory is not particularly attractive.

On the other hand, we can simply define the model we would like to study by
writing down the field theory
\begin{equation}
S=\int d^{2}x\left[ {\sum_{a=1}^{n}(}|D\psi _{a}|^{2}+m^{2}|\psi
_{a}|^{2})-\rho e^{-{\frac{1}{{z}}}{}{\sum_{a=1}^{n}}|\psi
_{a}|^{2}}\right]   \label{Smag}
\end{equation}
with the covariant derivative $D_{j}=\partial _{j}-iA_{j}$. Now the inverse $%
f(x,y)$ of $f^{-1}(x,y)=(-D^{2}+m^{2})\delta (x-y)$ does not have a
particularly simply form. The second form, Eq. (\ref{Smag}) arises
from the magnetic version of model II.\cite{Affleck}

Thus, we can define two different classes of models to study density of
states and localization in the presence of a magnetic field. We can either
have a simple $f(x,y)$ or a simple $f^{-1}(x,y).$ 

Suppose we want to calculate the density of states for $\rho $
large. Doing a high density expansion of the Green's function, we have
\begin{equation}
G(z)=\sum_{n=0}^{\infty }{\frac{1}{z^{n+1}}}\rho ^{n}\int
dx_{1}dx_{2}...dx_{n}f{\left( x_{1}-x_{2}\right) }...f{\left(
x_{n-1}-x_{n}\right) }e^{{\frac{i}{2}}B\left[ \left( \vec{x_{1}}\times \vec{
x_{2}}\right) +\left( \vec{x_{2}}\times \vec{x_{3}}\right) +...+\left( \vec{
x_{n}}\times \vec{x_{1}}\right) \right] }
\end{equation}
After Fourier transforming
\begin{equation}
f(x)=\int \frac{d^{2}k}{(2\pi )^{2}}e^{i\vec{k}\vec{x}}f(k)
\end{equation}
we can do the Gaussian integrals over $x$ to obtain
\begin{equation}
G(z)=\sum_{n=0}^{\infty }{\frac{1}{z^{n+1}}}\rho ^{n}\int \frac{%
d^{2}k_{1} }{(2\pi )^{2}}...\frac{d^{2}k_{n}}{(2\pi )^{2}}f\left(
k_{1}\right) ...f{\ \left( k_{n}\right) }e^{\frac{i}{2Bn}\sum_{q}\frac{1}{%
2cosq}\left( \vec{T} \times \vec{T}^{*}\right) }
\end{equation}
where
\begin{equation}
\vec{T}=\sum_{j}e^{iqj}\vec{p}_{j}
\end{equation}
where $\vec{p}_{j}=\vec{k}_{j}-\vec{k}_{j-1}.$ We were not able to evaluate $%
G(z)$ but in principle it might be possible for a Gaussian $f(k)=e^{-ak^{2}}$
.

In the $B\rightarrow 0$ limit the stationary phase requirement forces all
the $\vec{k}$'s to be equal and so we recover the appropriate zero magnetic
field result in Ref. (\onlinecite{MPZ}). In the opposite $B\rightarrow
\infty $ limit we drop the exponential and obtain to leading order $G(z)=
\frac{1}{z-\rho f(x=0)},$ a result as expected. It simply says that as the
Larmor radius goes to zero, each point in our random collection of points is
its own universe. Again, in principle, the corrections in powers of $1/B$
could be worked out.

\acknowledgments One of us (AZ) would like to thank Th. Nieuwhenhuizen for
helpful discussions and for hospitality at the University of Amsterdam. The
research of IA is supported in part by NSERC of Canada and by the NSF grant
PHY94-07194 and that of AZ is supported in part by the NSF grants
NSF89-04035 and PHY94-07194.

\end{document}